\documentclass{elsart}
\usepackage{natbib}
\usepackage{graphicx}
\usepackage{dcolumn}
\usepackage{bm}
\usepackage{epsf}
\usepackage{amssymb}
\usepackage{array}

\newcommand{\cuo}{$\mathrm{CuO_2}$}

\begin{document}
\begin{frontmatter}
\title{Anisotropic properties of spherical single crystals of La$_{\mathbf{1.8}}$Sr$_{\mathbf{0.2}}$CuO$_{\mathbf{4-\delta}}$ }

\author[UU,CMU]{A. Gardchareon}
\author[CMU]{N. Mangkorntong}
\author[UU]{D. H\'{e}risson }
\author[UU]{P. Nordblad}

\address[UU]{Department of Engineering Sciences, Uppsala
University, Box 534, SE-751 21 Uppsala, Sweden}
\address[CMU]{Faculty of Sciences, Chiang Mai University, Chiang
Mai 50200, Thailand}
\date{\today}

\begin{abstract}
Single crystals of $\mathrm{La_{1.8}Sr_{0.2}CuO_{4-\delta}}$ grown
by the travelling solvent floating zone method have been ground to
spherical shape for studies of anisotropic supercondcucting
properties by SQUID magnetometry. Here we report on magnetization
measurements parallel and perpendicular to the c-axis on two of
these crystals. At low enough temperatures and fields the spheres
are perfectly shielding (susceptibility -1.5 [SI]) and thus
magnetically isotropic. The field dependence of the critical
temperature, the transition width and the field expulsion well
below $T_c$ is qualitatively similar in both samples and is
detailed in the paper. The anisotropy of the first critical field
also shows similar behavior for the two crystals. Magnetization vs
field experiments reveal a large difference in the hysteresis
behavior along the two directions and an enhancement of the
critical current density through the \cuo-planes compared to the
in-plane current density at higher fields.
\end{abstract}

\begin{keyword}
Magnetic properties, Type II superconductor, High $T_c$ superconductor, Anisotropy, $\mathrm{La_{2-x}Sr_{x}CuO_{4-\delta}}$

\PACS 74.25.Qt \sep 74.72.Dn \sep 75.30.Gw
\end{keyword}

\maketitle

\end{frontmatter}

\section{Introduction}
$\mathrm{La_{2-x}Sr_{x}CuO_{4-\delta}}$ is a high-$T_c$
superconducting compound~\cite{Kishio1,Tarascon2,VanDover3} with a
crystal structure that contains only one $\mathrm{CuO_2}$ plane
per primitive cell. The superconductivity of the compound depends
strongly on the Sr concentration~\cite{Tarascon2,VanDover3} and
the physical properties of the system have been quite extensively
investigated~\cite{Iwasaki4,Sasagawa5,Tachiki6,Lake7,Yung8,Gaojie,Li,Duran2,Levy,Leylekian,Maletta,Decca,Shcherbakov,Leylekian2,Rodrguez,Levy2} Here we report  results from magnetization measurements parallel and
perpendicular to the c-axis of two spherical single crystal of
nominally the same composition
$\mathrm{La_{1.8}Sr_{0.2}CuO_{4-\delta}}$ (LSCO) still with
quantitatively dissimilar properties.

One purpose of the study is to investigate the anisotropy of
the magnetic response of an extreme type II superconducting system
without any disturbing geometrical anisotropy at the moment of
field penetration. Another objective of the study is to make
a detailed mapping of the superconductivity of
$\mathrm{La_{2-x}Sr_{x}CuO_{4-\delta}}$ as to the generality of
the magnetic response and the anisotropy for two samples of
nominally the same quality and composition.

\section{Experimental details}

  \subsection{Single Crystal Growth}

The single crystals of  $\mathrm{La_{2-x}Sr_{x}CuO_{4-\delta}}$
were prepared by means of the travelling solvent floating zone
(TSFZ) method~\cite{Tanaka9} using an optical image furnace of 4
cups type. The starting material was prepared by a solid state
reaction from high purity (4N) powders of $\mathrm{La_2O_3}$,
$\mathrm{ SrCO_3}$ and CuO, formed into a powder rod by means of
water press and sintered in air at $1260 ^o$C for 12 hr. To start
the crystal growth, a powder rod was hanged inside a fused quartz tube,
with the lower end of the rod slightly above a single crystal seed
and at the focal point of an optical image furnace. Oxygen at a
slightly higher than ambient pressure was admitted to flow through
the quartz tube. A melting zone was created at the lower end of
the rod by the heat generated from a set of four 300 watt halogen
lamps. In order to lower the melting point, a piece of $\mathrm{CuO}$, was
employed as solvent. The crystal seed was moved up and connected
to the melting zone of the powder rod to initiate the growth of
the single crystal. The powder rod and the single crystal seed
were rotated in opposite directions at a speed of 15 -- 30~rpm.
Then they were slowly moved downward together at a rate of about
0.8~mm/hr while the melting zone remained at the focus of the lamp
images causing the lower part of the rod to become a solid single
crystal. This process was carried out until a complete single
crystal rod had been formed. Typical dimensions of the single
crystal rods were 8 -- 10~cm of length and 4 -- 6~mm in diameter.
The crystal rods were always annealed in one atm of oxygen, at
$800 ^o$C for 240 hours prior to any characterization.

\begin{figure*}[htb]
\centering
\includegraphics[width=0.9\textwidth]{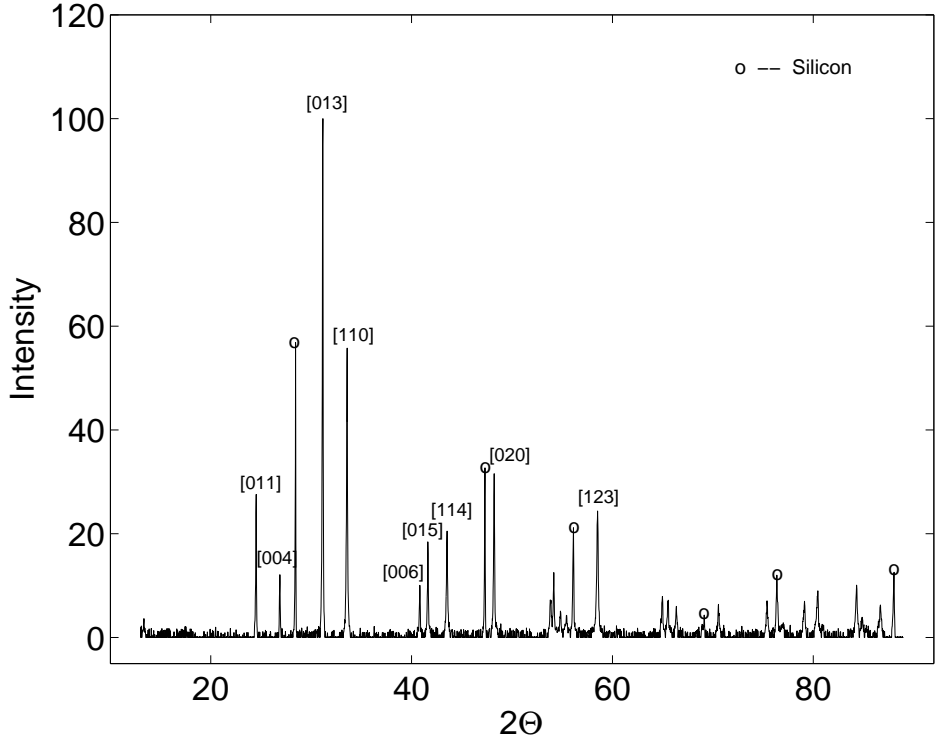}
\caption{A powder x-ray diffraction pattern of a LSCO sample. The strongest LSCO peaks are labelled
and the Si peaks marked by circles. $\lambda$ = 1.5406~\AA.}
\label{fig1}
\end{figure*}

A powder X-ray diffraction pattern of a sample extracted from one
of the LSCO crystals is  shown in Fig.~\ref{fig1}. It shows that
this sample is single phased without noticeable impurities (~
$\leq$ 0.5\%). Composition analyses of several single crystal
rods were carried out by means of an Electron Probe Microanalyzer
instrument (JEOL EPMA), which revealed that the composition of the
LSCO crystals was homogeneous and uniform throughout the length
of the rods. Single domain examination was done by a metallurgical
optical microscope and Laue photography. From these
characterization results the single crystals were found to be of
high quality. Some special properties of these LSCO crystals have
been studied with neutron scattering~\cite{Lake7}.
\\
Two single crystal rods of
$\mathrm{La_{1.8}Sr_{0.2}CuO_{4-\delta}}$ were selected
for further studies. It was found from powder x-ray diffraction
measurement that these two samples have tetragonal structure
with a = 3.7738$\pm$0.0001~\AA, c = 13.2467$\pm$0.0008~\AA$ $ for the first crystal
(sample 1) and a = 3.7719$\pm$0.0001~\AA, c = 13.2528$\pm$0.0007~\AA$ $ for the
second crystal (sample 2). Comparing these parameters to the relation between
 $\mathrm{Sr}$ content x and the lattice constants at room temperature reported by
 Takagi et al.\cite{Takagi10}, the $\mathrm{Sr}$ content x for our samples was
 confirmed to be close to the nominal x $\approx$ 0.2. The transition
 temperatures ($T_c$) of the two crystals were found to be
  30.5~K and 33.5~K for sample 1 and sample 2, respectively.

\subsection{Experimental procedures}

Spherical single crystal samples were ground from the two single
crystal rods. Pieces of about 4~mm of length were cut from the
rods by a diamond saw. The pieces were first hand ground and then
ground to good spherical shape in a special sphere grinding tool.
The diameter of the two spheres employed in this study is about
2~mm, and the mass of the crystals are 28.90~mg (sample 1) and
30.80~mg (sample 2), respectively. The c-axis directions of the
spheres were determined (and clearly marked) by using the single
crystal diffractometer (SXD) at the Studsvik Neutron Research
Laboratory, Sweden.

A commercial SQUID magnetometer (Quantum Design MPMSXL) with a 5~T
magnet was used to measure the magnetic properties of the samples.
The temperature dependence of zero-field-cooled (ZFC),
field-cooled (FC) and thermoremanent (TRM) magnetization was
measured in the temperature range 5 -- 40~K with the magnetic field
applied parallel and perpendicular to the c-axis of the samples.
The measurements were performed in different applied magnetic
fields ranging between 0.08 and 40~kA/m. The field dependence of the
magnetization (full hysteresis curves) of the two samples in fields up to $4\times 10^3$~kA/m
 was measured at different constant temperatures
and with the field applied parallel and perpendicular to the
c-axis of the samples.

The zero-field-cooled magnetization ($M_{\rm ZFC}$) is obtained by
cooling the sample to a temperature below the transition
temperature $T_c$ in the absence of the field, turning on an
applied magnetic field, and measuring the magnetization in this
field as the sample warms through $T_c$. The field-cooled-cooling
magnetization ($M_{\rm FC}$) is obtained by cooling the sample in an
applied magnetic field and measuring the magnetization in this
field as the sample is cooled. All FC results of this paper are recorded
on cooling, however, for comparision, some FC-magnetization were also
recorded on re-heating the sample; these data did not deviate significantly
from the cooling results. The thermo-remanent-magnetization ($M_{\rm TRM}$) is obtained
by turning off the field after field-cooling and measure the magnetization as the sample is reheated through $T_c$.
\\

\section{Results and discussion}

\subsection{Temperature dependence of the magnetization}

The crystals were first investigated by recording the temperature
dependence of the magnetization at different applied fields along
the two principal axes of the samples using the ZFC, FC and TRM
protocols as described above. From the low field $M$ vs $T$ curves the
transition temperatures of the two samples were determined to be 30.5~K and 33.5~K for sample 1 and 2,
respectively.(  $T_c$ is here defined from the temperature for the onset of a
diamagnetic signal in the ZFC and FC magnetization curves.)

\begin{figure}[ht]
\centering
\includegraphics[width=0.75\textwidth]{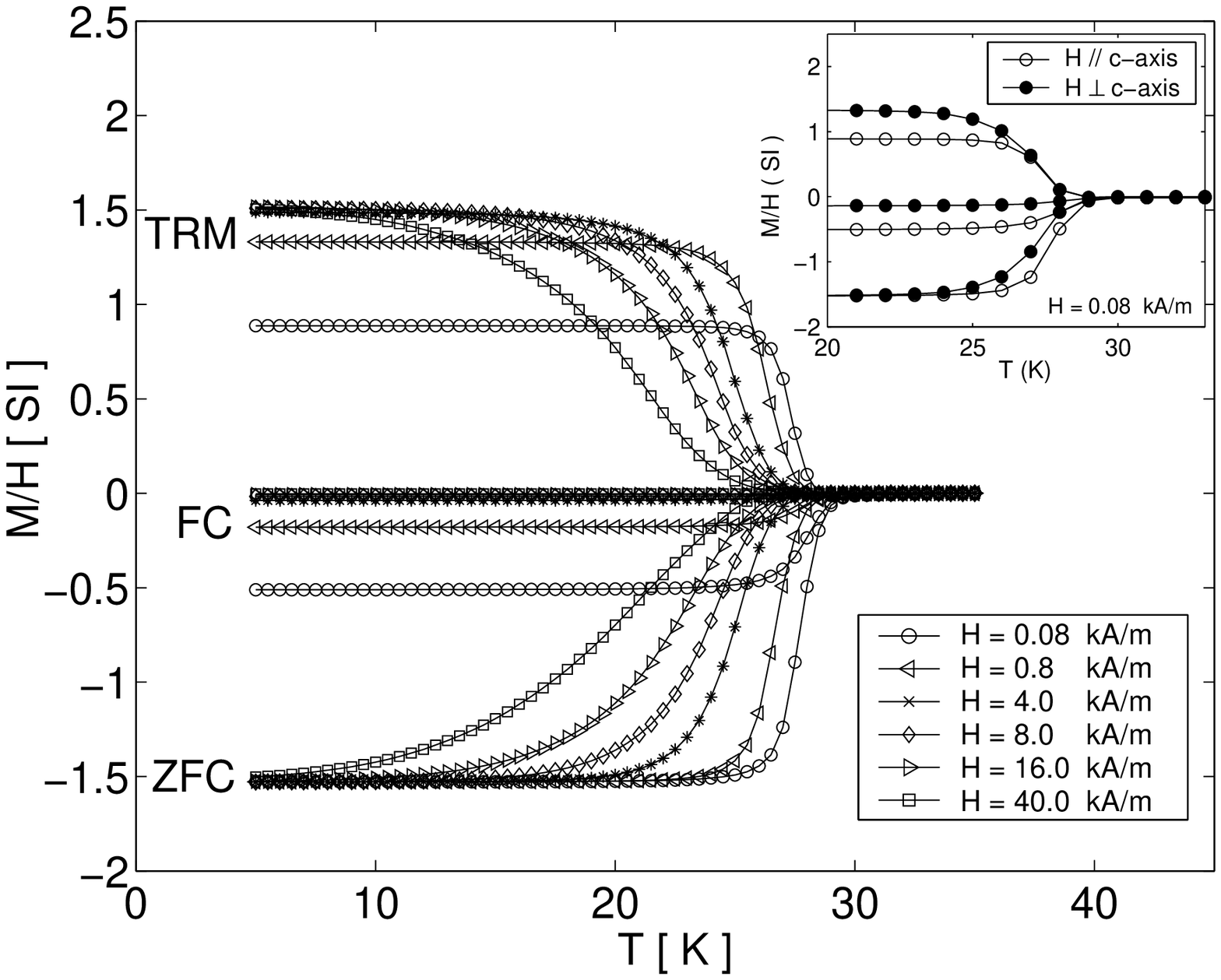}
\includegraphics[width=0.75\textwidth]{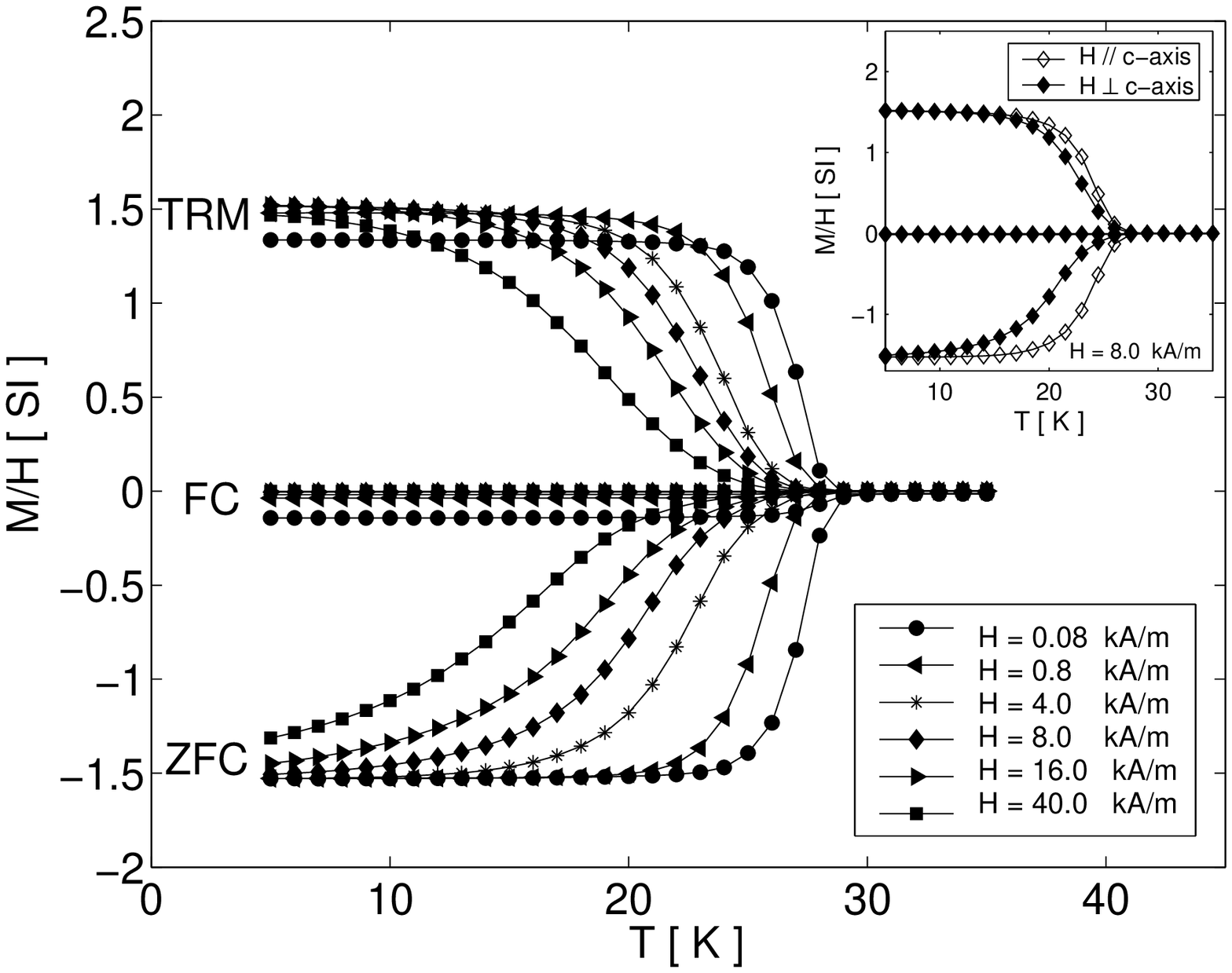}

\caption{Temperature dependence of the ZFC, FC and TRM
``susceptibility'' (M/H, [SI]) for the field parallel (\textsc{Top}) or perpendicular (\textsc{Bottom}) to the c-axis
of sample 1, $H$ = 0.08, 0.8, 4.0, 8.0, 16.0 and 40.0~kA/m. The insets show
the temperature dependence of the ZFC, FC and TRM ``susceptibility''
for the field parallel and perpendicular to the c-axis, for $H$ = 0.08~kA/m (\textsc{Top}) and $H$ = 8.0~kA/m (\textsc{Bottom}).}
\label{fig2}
\end{figure}

Fig.~\ref{fig2} \textsc{Top} shows $M/H$ vs $T$ at different applied fields
for sample 1 with the field applied along the c-axis. It should be
noted that the value of $M_{\rm ZFC}/H$ = -1.5 at low temperatures
corresponds to the perfect shielding value for a superconductor of
spherical shape ($\chi = -\frac{1}{1-N}$, where  $N = \frac{1}{3}$ is the demagnetising factor,
leads to $\chi=-1.5$). As is seen from the figure the
field expulsion ($M_{\rm FC}/H$), on the other hand, is quite weak and
rapidly decreasing with increasing field.

Fig.~\ref{fig2} \textsc{Bottom} shows $M/H$ vs $T$ at different applied fields for
sample 1 with the field applied perpendicular to the c-axis. The
low temperature value of $M_{\rm ZFC}/H$ = -1.5, corresponds to
perfect shielding as when the field is applied parallel to the
c-axis. The field expulsion in this case is even weaker than when
the field is applied parallel to the c-axis.

The insets of Fig.~\ref{fig2} give a comparison
between the behavior of $M/H$ for $H$ parallel and perpendicular
to the c-axis at a weak field, $H$ = 0.08 kA/m (\textsc{Top} inset) and a somewhat stronger field $H$ = 8.0 kA/m  (\textsc{Bottom} inset). It is clearly seen from the figure that $M/H$
is more strongly suppressed by the field when the field is applied
perpendicular to the c-axis, i.e, when the induced supercurrents
must travel through the areas linking the superconducting
\cuo -planes.

The corresponding curves for sample~2 give a qualitatively similar
picture as Fig.~\ref{fig2} for sample~1, however
with some quantitative differences, which we will discuss below.

\begin{figure}[ht]
\centering
\includegraphics[width=0.9\textwidth]{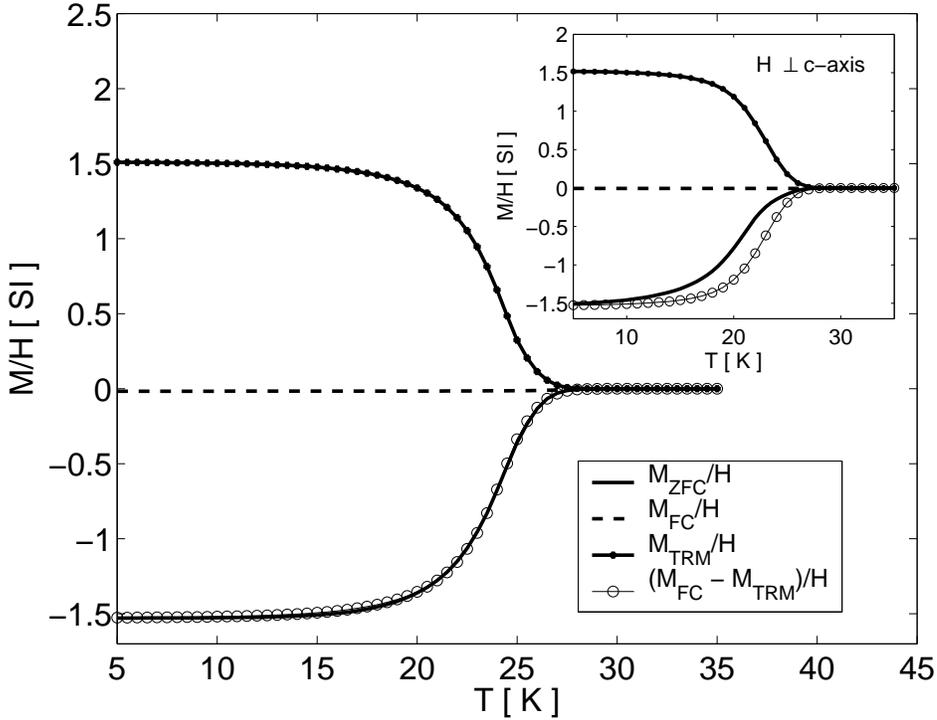}
\caption{Temperature dependence of the ZFC, FC and TRM
``susceptibility'' for the field parallel to the c-axis of sample~1 ($H$ = 8.0~kA/m), the open squares represent $M_{\rm FC}-M_{\rm TRM}$,
which is indistinguishable from $M_{\rm ZFC}$ and confirms the
validity of the principle of superposition. The inset shows the
corresponding curves when the 8.0~kA/m field is applied
perpendicular to the c-axis and shows that the principal of
superposition is no longer valid.}
\label{fig3}
\end{figure}

The fundamental relation $M_{\rm ZFC}=M_{\rm FC}-M_{\rm TRM}$, which
is based upon the principle of superposition and only valid for a
system that yields linear response to a field change, is found to
apply for low fields, but to be violated for higher fields. The
relation is valid up to higher magnetic fields applied parallel to
the c-axis than perpendicular to the c-axis as can be seen in
Fig.~\ref{fig3}, where $M_{\rm ZFC}=M_{\rm FC}-M_{\rm TRM}$ data for
an applied field of 8.0~kA/m applied parallel (main frame) and
perpendicular to the c-axis (inset) are shown for sample~1. The
validity of the relation is only certified when the sample is able
to fully shield a field change, i.e., when there is no further
field penetration. Additionally, the trapped field from the field
cooled process must remain trapped when the field is removed for the
measurement of the TRM state. The two samples show very similar
behavior as to the applicability of the principle of
superposition.

\begin{figure}[ht]
\centering
\includegraphics[width=0.9\textwidth]{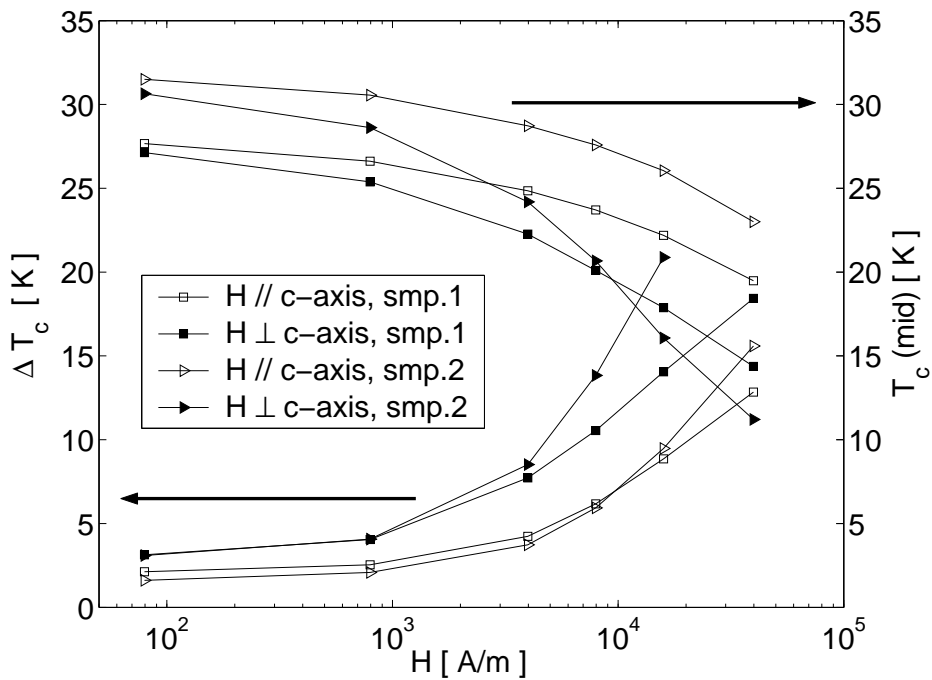}
\caption{Field dependence (logarithmic scale) of $\Delta T_c$
and $T_{c}$(mid) for the magnetic field applied parallel and
perpendicular to the c-axis for sample 1 and sample 2.}
\label{fig4}
\end{figure}

An important consequence of an increasing applied magnetic field on a
high temperature superconductor is that the superconducting temperature is lowered and that the
width of the transition increases.   
Fig.~\ref{fig4} illustrates the increase of the width of the transition $\Delta T_c$, defined from the difference between the temperatures for which the shielding has reached 10\% and  90\% of the full shielding level, and the decrease of $T_c$(mid), here defined through the temperature where half of full shielding is measured. The change of the parameters is plotted vs. applied magnetic field for both samples and with the field applied
parallel and perpendicular to the c-axis. The figure shows that
the increase of $\Delta T_c$ and the decrease of $T_c$(mid) is much
stronger for magnetic fields perpendicular to the c-axis than for
magnetic fields parallel to the c-axis and again that the behavior
of the two samples is similar. However, a somewhat stronger field dependence for
the sample with the higher $T_c$ (sample 2) can be noticed. It should be mentioned that the decrease of $T_c$(mid) with the field is stronger than the decrease of $T_c$ onset.

\begin{figure}[ht]
\centering
\includegraphics[width=0.9\textwidth]{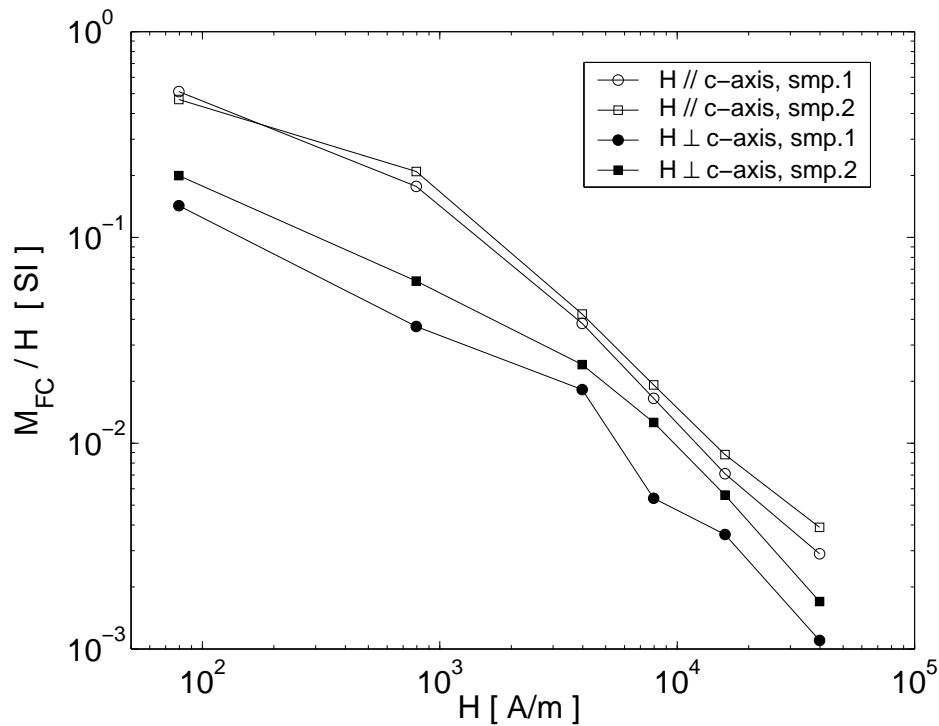}
\caption{Field dependence of the low temperature field cooled
``susceptibility'' $(M_{\rm FC})/H$ (temperature independent at low
enough temperatures), when the field is applied parallel and
perpendicular to the c-axis of sample 1 and sample 2.}
\label{fig5}
\end{figure}

Fig.~\ref{fig5} shows the field dependence of the magnitude of the
field expulsion ($M_{\rm FC}$) of sample 1 and 2 for applied magnetic
fields parallel and perpendicular to the c-axis. The magnitude of
$M_{\rm FC}$ for applied magnetic fields parallel to the c-axis is
higher than for fields applied perpendicular to the c-axis. The
field expulsion is quit weak, at a field of 0.08~kA/m it amounts to
$\sim$~30\% of complete field expulsion for 
fields parallel to the c-axis, but only $\sim$~10\% for fields
perpendicular to the c-axis for both samples and it is rapidly
decreasing with increasing field. The small value of the flux
expulsion even at quite weak fields implies that the flux lines are
always strongly trapped in these samples.

\subsection{Field dependence of the magnetization}

\begin{figure}[ht]
\centering
\includegraphics[width=0.75\textwidth]{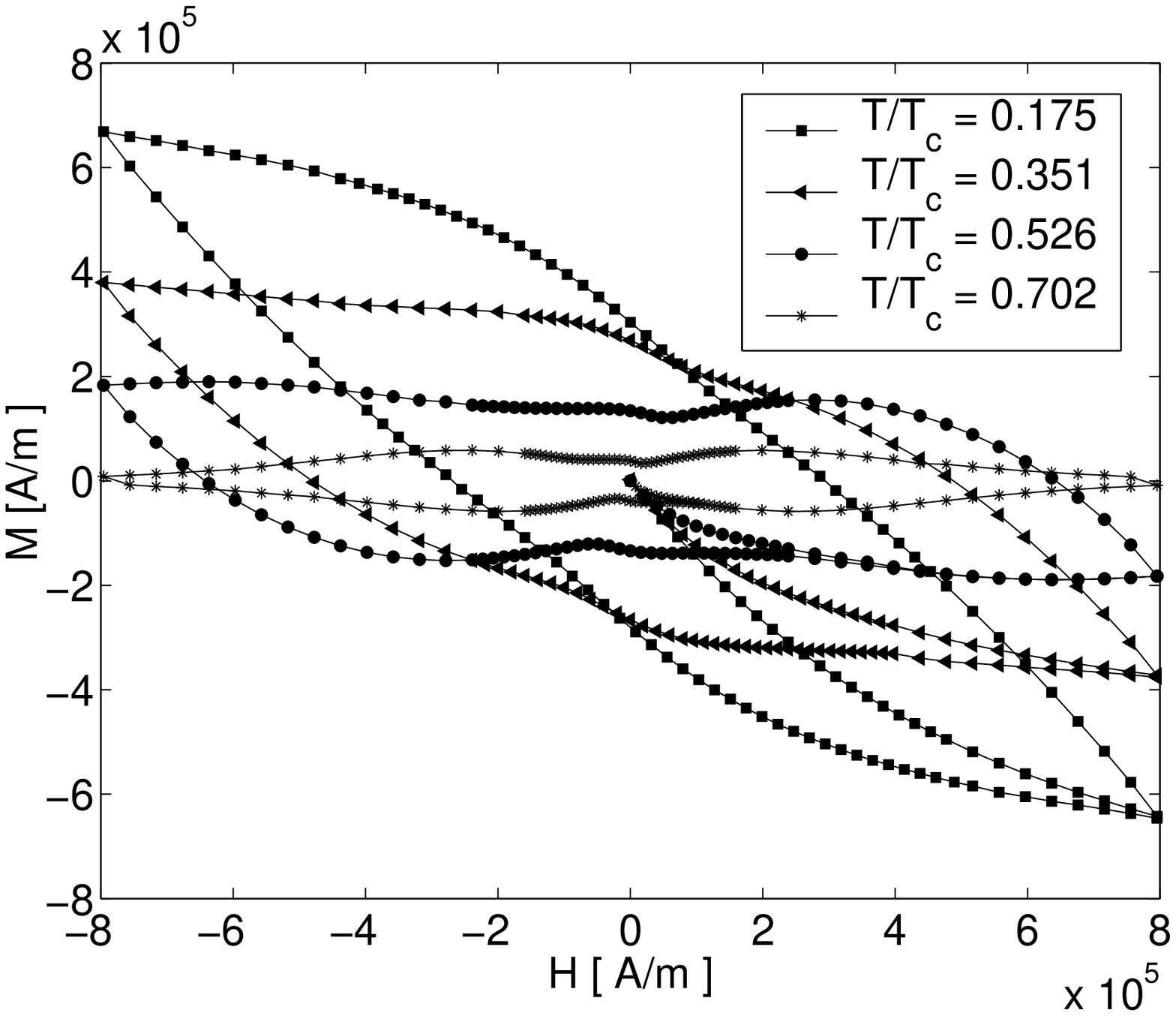}
\includegraphics[width=0.75\textwidth]{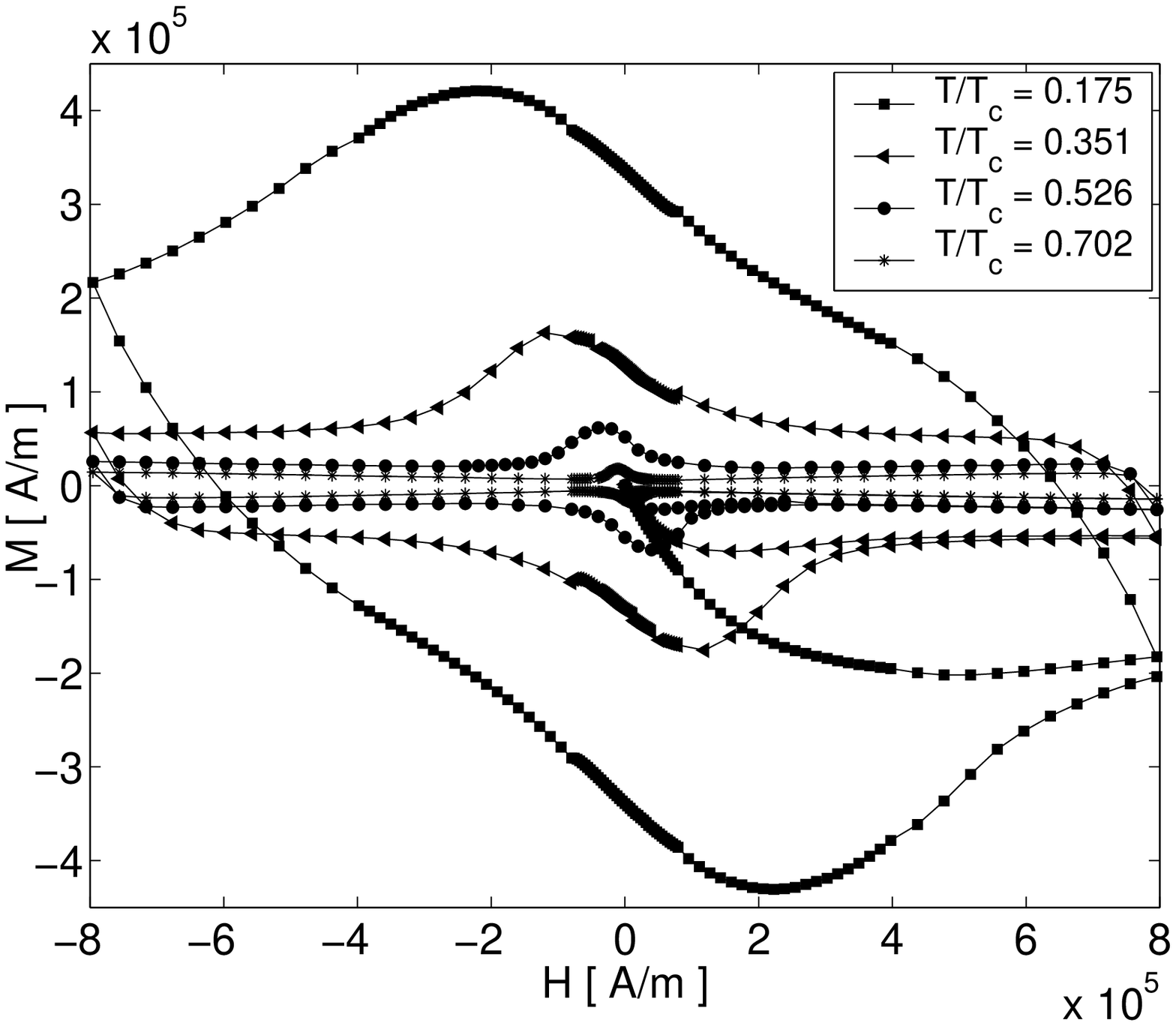}
\caption{Field dependence of magnetization at $T/T_c=0.175$,
0.351, 0.526 and 0.702 for magnetic fields applied parallel (\textsc{Top}) and perpendicular (\textsc{Bottom}) to the
c-axis of sample 1.}
\label{fig6}
\end{figure}

The upper graph of Fig.~\ref{fig6} shows the field dependence of the magnetization in
the field range  -800~kA/m to 800~kA/m at four different temperatures in the
range $T/T_c = 0.175$ -- 0.702, when the field is applied parallel
to the c-axis for sample 1 (sample 2 shows a similar behavior). It
is seen that the hysteresis loop becomes more narrow with
increasing temperature, a direct reflection of the decrease of the
first critical field ($H_{c1}$) and the critical current density ($J_c$) of
the samples with increasing temperature. One also notices that the
field for full flux penetration (the minimum in the virgin M vs H curve)
is not reached in these experiments at the two lower temperatures.

The lower part of Fig.~\ref{fig6} shows the corresponding $M$ vs $H$ curves when the
field is applied perpendicular to the c-axis of sample 1 (a
similar behavior is observed for sample 2). The width of the
hysteresis loop is significantly smaller than when magnetic field
is applied parallel to the c-axis and the field for full flux
penetration is reached at all the measured temperatures.

Some features of these hysteresis loops are worth to note: The
initial slope of all curves corresponds to perfect shielding i.e.
M/H = -1.5, the curves start to deviate from this straight line at a
quite well defined field value, the effective first critical field
($H_{c1eff}=H_{c1}/1.5$), and the continued M vs H loop for the
different directions imply that there are strongly temperature and
directional dependent pinning forces and critical currents in the
samples.

\begin{figure}[ht]
\centering
\includegraphics[width=0.9\textwidth]{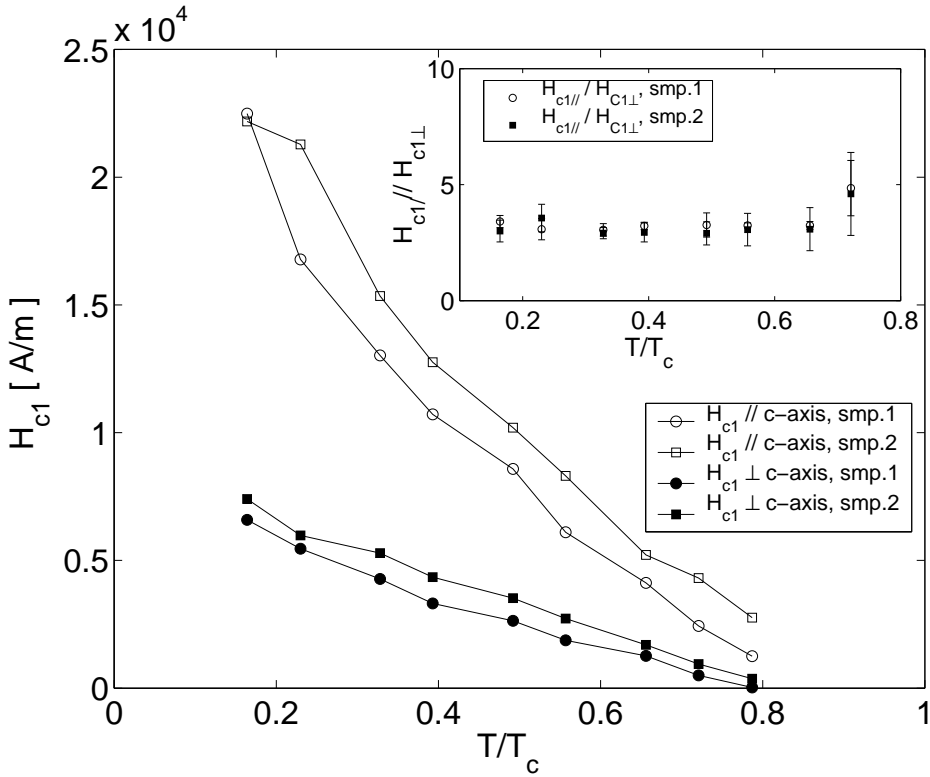}
\caption{H-T phase diagram showing $H_{c1}$ vs $T/T_c$ for the two
samples both parallel and perpendicular to the c-axis. The inset
shows the calculated anisotropy of $H_{c1}$ ($H_{c1\parallel}/H_{c1\perp}$) for the two samples.}
\label{fig7}
\end{figure}

Using the low field initial curves of Fig.~\ref{fig6}
and the corresponding curves for sample~2, we can
derive estimates of the first critical field of the samples. In
Fig.~\ref{fig7}, $H_{c1}$ is plotted vs temperature ($T/T_c$) for both
field directions and for the two samples. In the inset the
anisotropy of the first critical field defined from
($H_{c1\parallel}/H_{c1\perp}$) is shown.
The first critical field $H_{c1}$ is found to increase almost linearly on decreasing temperature and to show an anisotropy independent of temperature as shown by the inset of Fig.~\ref{fig7}. The average value of the $H_{c1}$-anisotropy, weighted by errorbars, is found to be $3.1\pm 0.1$. The measured data of the field dependence of $H_{c1}$ is somewhat scattered, but it is a great advantage to use a spherical single crystal to study the behaviour of the first critical field of a highly anisotropic superconductor, since
geometrical factors are effectively avoided.

In a standard Ginzburg-Landau Model of supraconductivity~\cite{GLM,VortexState00}, $H_{c1}$ is related to the order parameter coherence length $\xi$ and the field penetration depth $\lambda$ with a good approximation through~\cite{HuPRB6}:
$$\mu_0 H_{c1}=\frac{\Phi_0}{4\pi \lambda ^2}\left( \ln(\lambda / \xi) + 1/2\right)$$
where $\Phi_0=h/2e$ is the quantum of flux. In the case of an anisotropic material, this relation remains valid for fields applied along the principal axis. For an applied field along the c-axis, a good approximation of the right hand side parenthesis of the equation can be obtained using the $\xi$ measurements on LSCO samples of the same composition~\cite{Li}.
Using an extrapolation of the $H_c1$ measurements to $T=0$, this gives for our samples a field penetration depth $\lambda_{\parallel}(T=0)$ of about 1300~\AA.

 In this context, the anisotropy is described by the ratio $\alpha={\lambda_{\perp}\over \lambda_{\parallel}}={\xi_{\parallel}\over \xi_{\perp}}$~\cite{Klemm00}, and can be evaluated to $2.00\pm 0.05$, in agreement with value reported by Zaleski and Klamut~\cite{Zaleski00}. In some contrast; $\lambda_{\parallel}$ at $T=0$ was found to be about 2500~\AA~\cite{Li,Klemm00}.

\begin{figure}[htb]
\centering
\includegraphics[width=0.75\textwidth]{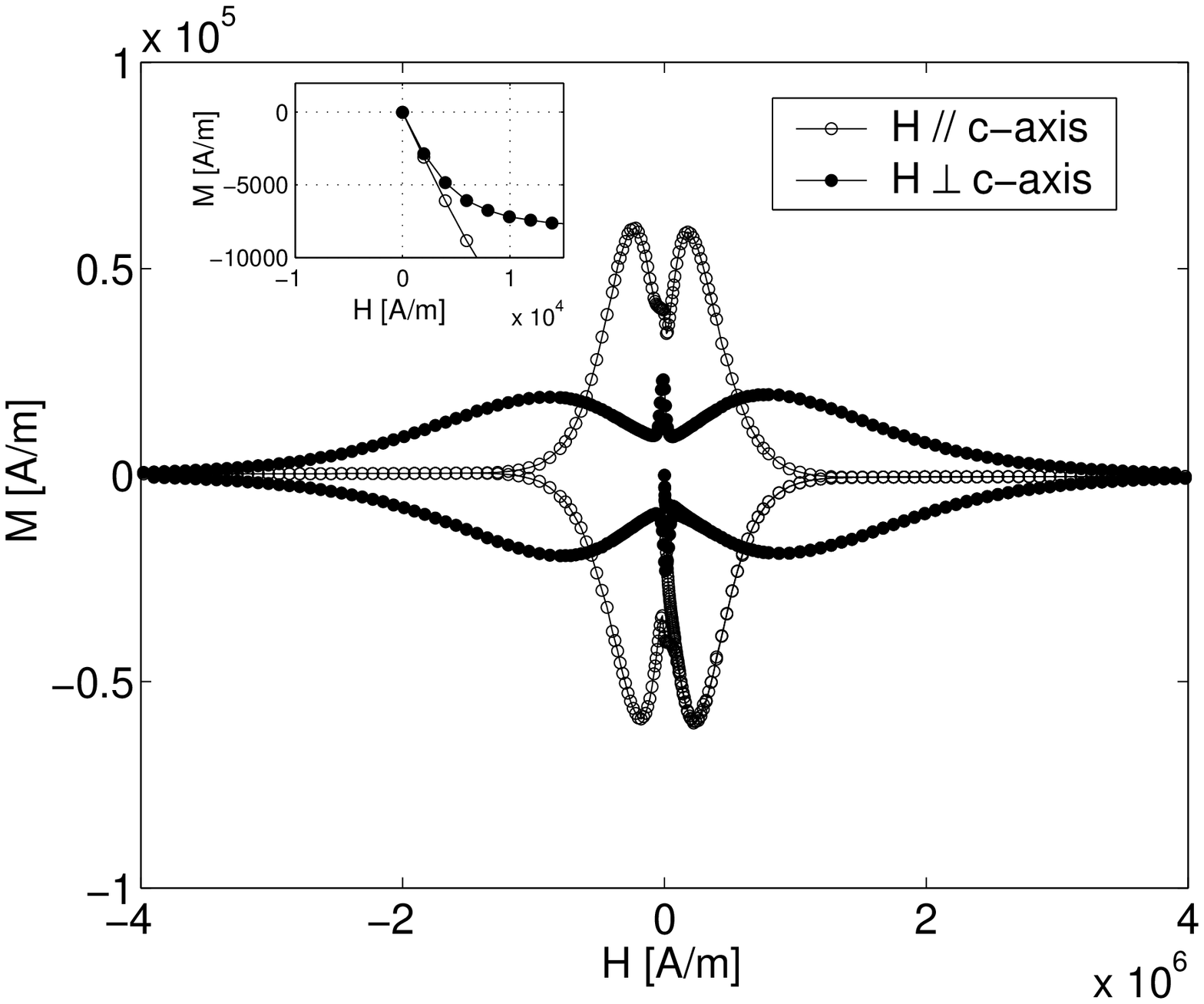}
\includegraphics[width=0.75\textwidth]{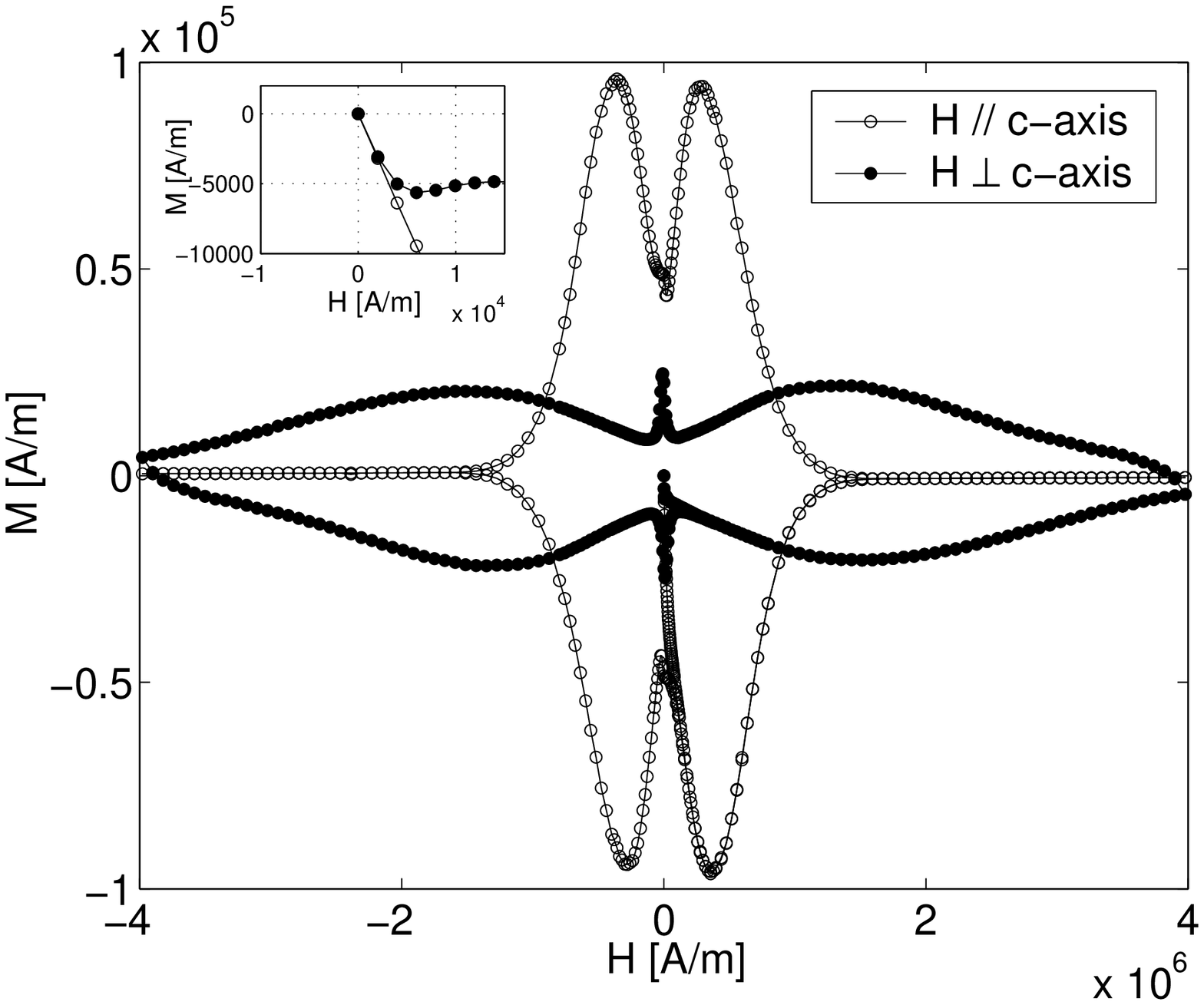}
\caption{Field dependence of the magnetization at $T/T_c = 0.702$  when
the field was applied parallel to the c-axis (open circles) and
perpendicular to the c-axis (solid circles). The inset shows the initial slope for the field dependence of magnetization when magnetic field is applied parallel and perpendicular to the c-axis.
\textsc{Top}: measured on sample~1; \textsc{Bottom}: measured on sample~2.}
\label{fig8}
\end{figure}

Fig.~\ref{fig8} shows the field dependence of the
magnetization at $T/T_c = 0.702$ for both samples,
when the field range is extended, $-4\times 10^3$ kA/m to $4\times 10^3$ kA/m, for applied magnetic
fields parallel and perpendicular to the c-axis. A qualitatively
similar behavior is seen for the two samples, however with a
significantly wider hysteresis loop (higher critical currents) for
the sample with the higher critical temperature (sample 2). It is
striking to note that there is a crossover field, where the
hysteresis loops for the field parallel to the c-axis becomes less
wide than for the perpendicular direction. The hysteresis loops in
the two directions do in fact show quite different behavior: The
``parallel'' curves exhibit a minimum at $H= 0$ followed by a quite sharp
maximum after which the magnetization rapidly decreases to become
very small at higher fields ($>800$~kA/m). The perpendicular hysteresis
loops on the other hand shows a sharp maximum at $H = 0$ followed by a
narrow minimum and a broad maximum at a larger field ($\approx 800$~kA/m)
 and a slow decrease of the magnetization as the field is further increased. A broad secondary maximum in the magnetization curve
have been observed for many different HTS compounds and certain modele have been proposed to explain this (anomalous) behaviour.\cite{Bugoslavsky2}

As expected for a cuprate superconductor, the anisotropy of the magnetic response is quite remarkable. The system shows a higher first critical field and higher critical currents when the shielding currents can travel only within the \cuo-plane than when they are forced to penetrate the inter-layers. However, at higher magnetic fields, the width of the hysteresis loop, and thus the observed critical currents ($J_c\propto \Delta M$ ) and the pinning forces become exceedingly larger in the perpendicular than in the parallel configuration. In table~\ref{tab:currents} the critical current density with the field applied parallel to the c-axis of the sample ($J_{c//}$) and the critical current density with the field perpendicular to the c-axis of the sample ($J_{c\perp}$) are reported. They have been calculated from the magnetization for a full critical state established in a spherical shape sample $\Delta M = 3\pi J_c a/16$, where $a$ is the sphere radius~\cite{ConcEncyP99}.
\begin{table}
\centering
\begin{tabular}{|l|c|c|c|c|}

\hline
 	&\multicolumn{2}{c}{$H_{m\parallel}$}\vline&\multicolumn{2}{c}{$H_{m\perp}$}\vline\\
\cline{2-5}
	&$J_{c\parallel}$ (A/m$^2$)&$J_{c\perp}$ (A/m$^2$)&$J_{c\parallel}$ (A/m$^2$)&$J_{c\perp}$ (A/m$^2$)\\
\hline
Sample 1&$2.0\times10^8$&$4.0\times10^7$&$1.5\times10^7$&$6.5\times10^7$\\
Sample 2&$3.2\times10^8$&$4.0\times10^7$&$6\times10^5$&$7\times10^7$\\
\hline
\end{tabular}
\caption{Critical currents evaluated at the different maxima of Fig.~\ref{fig8}, i.e. $H_{m\parallel}\approx 2.2\times 10^5$~A/m and $H_{m\perp}\approx 8.7\times 10^5$~A/m for sample~1 and $H_{m\parallel}\approx 3.5\times 10^5$~A/m and $H_{m\perp}\approx 1.5\times 10^6$~A/m for sample~2.}
\label{tab:currents}
\end{table}
The shape of the curve and the deduced critical current values show a strong field dependence, which is not taken into account in the critical state model used to evaluated $J_c$. However, the results imply a much stronger field dependence of the pinning forces and the flux dynamics when the flux lines penetrate the \cuo-{planes} perpendicularly than when the flux lines are trapped in between the \cuo-planes.
For the field parallel to the c-axis the current density on the timescale of our experiments, becomes lower than the resolution of our setup, whereas it maintains a considerable magnitude at the corresponding field in the perpendicular direction.

\section{Conclusions}

      The magnetic response of two spherical single crystals of LSCO of nominally the same composition, has been found to show qualitatively similar behaviour, although they have somewhat different critical temperatures. At low fields and low temperatures the crystals are perfectly shielding but the field expulsion is week. The detailed behaviour on approaching the critical temperature shows some significant differences. The sample with the higher critical temperature (sample 2) has a somewhat sharper transition ($\Delta T_c$ = 2.6~K and 2.1~K respectively). On the other hand, the sample with the lower critical temperature shows a wider hysteresis loop in both field directions at the same relative temperature.
The anisotropy of the samples can be described by two main features. The first critical field $H_{c1}$ is found to increase according to an H-dependence on decreasing temperature and to show an anisotropy of about 3.1 independent of temperature. In an anisotropic Ginzburg-Landau Model, it corresponds to an anisotropy of the order parameter coherence length ${\xi_{\parallel}\over \xi_{\perp}}=2.0$. The second aspect concerns irreversibilities. The layered structure of the supraconductivity affects the pinning of the vortices differently in the parallel and perpendicular directions. The critical currents are found to have a stronger field dependence for fields applied along the c-axis.

\section*{acknowledgements} P. Svedlindh is acknowledged for helpful discussions, H. Takagi for providing the crystal growth facilities at Takagi Lab., Univ. of Tokyo, P. Mangkorntong for crystal growth helping, H\aa{}kan Rundl\"of for the SXD measurements at Studsvik Neutron Research Laboratory and Nils Olov Ersson for the X-rays measurements.

Financial support from the Swedish Research
Council (VR) is acknowledged. A.G. acknowledges the International Science Program of Uppsala
University for administrating; the Ministry of Science,
Technology and Environment of Thailand for financing a fellowship
for research in Sweden.
D.H. acknowledges the support of the \textsc{Dyglagemem} European Research Training Network.

\bibliographystyle{elsart-num}
\bibliography{biblio}

\begin{thebibliography}{10}
\expandafter\ifx\csname url\endcsname\relax
  \def\url#1{\texttt{#1}}\fi
\expandafter\ifx\csname urlprefix\endcsname\relax\def\urlprefix{URL }\fi

\bibitem{Kishio1}
K.~Kishio, K.~Kitizawa, S.~Kanbe, I.~Yasuda, N.~Sugii, H.~Takagi, S.~I. Uchida,
  K.~Fueki, S.~Tanaka, New {H}igh {T}emperature {S}uperconducting {O}xide.
  ({L}a$_{1-x}${S}r$_x$)$_2${C}u{O}$_4$ and
  ({L}a$_{1-x}${C}a$_x$)$_2${C}u{O}$_4$, Chem. Lett. 2 (1987) 429--432.

\bibitem{Tarascon2}
J.~M. Tarascon, L.~H. Green, W.~R. McKinnon, G.~W. Hull, T.~H. Geballe,
  Superconductivity at 40 {K} in the oxygen-defect perovskites.
  {L}a$_{2-x}${S}r$_x${C}u{O}$_{4-y}$, Science 235 (1987) 1373--1376.

\bibitem{VanDover3}
R.~B. van Dover, R.~J. Cava, B.~Batlogg, E.~A. Rietman, Composition-dependent
  superconductivity in $\mathrm{La_{2-x}Sr_xCuO_{4- \delta}}$, Phys. Rev. B 35
  (1987) 5337--5339.

\bibitem{Iwasaki4}
H.~Iwasaki, F.~Matsuoka, K.~Tanigawa, Magnetization peak around {H}//a axis in
  {L}a$_{2-x}${S}r$_x${C}u{O}$_4$ single crystals with different anisotropy,
  Phys. Rev. B 59 (1999) 14624--14629.

\bibitem{Sasagawa5}
T.~Sasagawa, Y.~Togawa, J.~Shimoyama, A.~Kapitulnik, K.~Kitazawa, K.~Kishio,
  Magnetization and resistivity measurements of the first-order vortex phase
  transition in $\mathrm{(La_{1-x}Sr_x)_2CuO_4}$, Phys. Rev. B 61 (2000) 1610.

\bibitem{Tachiki6}
T.~Tachiki, K.~Nakajima, T.~Yamashita, I.~Tanaka, H.~Kojima, Flux {F}low of
  {L}a-{S}r-{C}u-{O} {S}ingle {C}rystals, IEEE. Trans. Appl. Supercon. 9 (1999)
  2191.

\bibitem{Lake7}
B.~Lake, H.~M. Ronnow, N.~Chistensen, G.~Aeppli, LefmannK., D.~F. McMorrow,
  P.~Vorderwisch, P.~Smeibidl, N.~Mangkorntong, T.~Sasagawa, M.~Nohara,
  H.~Takagi, T.~E. Mason, Antiferromagnetic order induced by an applied
  magnetic field in a high-temperature superconductor, Nature 415 (2002) 299.

\bibitem{Yung8}
Y.~M. Huh, D.~K. Finnemore, Vortex fluctuations in superconducting
  $\mathrm{La_{2-x}Sr_xCuO_{4+\delta}}$, Phys. Rev. B. 65 (2002) 092506.

\bibitem{Gaojie}
X.~Gaojie, P.~Qirong, Z.~Zengming, D.~Zejun, Effect of {O}xygen-{D}eficiency of
  $\mathrm{CuO_2}$ {P}lane on the {S}tructure, {M}agnetism, and {T}ransport
  {P}roperties, Journal of Superconductivity 14 (2001) 509--517.

\bibitem{Li}
Q.~Li, M.~Suenaga, Reversible magnetic properties of
  ({L}a$_{1-x}${S}rx)$_2${C}u{O}$_4$ single crystals with 0.05 $\leqslant$ x
  $\leqslant$ 0.10, Phys. Rev. B 47 (1994) 11384.

\bibitem{Duran2}
C.~A. Dur\'an, P.~L. Gammel, R.~Wolfe, V.~J. Fratello, D.~J. Bishop, Direct
  magneto-optical measurements of anisotropic critical currents in
  ({L}a$_{1-x}${S}rx)$_2${C}u{O}$_4$ single crystals, Phys. Rev. B 49 (1994)
  3608--3612.

\bibitem{Levy}
P.~Levy, H.~Ferrari, C.~Acha, V.~Bekeris, Irreversibility effects in
  polycrystalline high-{T}$_c$ superconductors studied by {AC} susceptibility,
  Physica C 222 (1994) 212--218.

\bibitem{Leylekian}
L.~Leylekian, M.~Ocio, M.~V. Feigelman, L.~B. Ioffe, Glassy superconductor
  state in $\mathrm{La_{1.8}Sr_{.2}CuO_4}$ ceramics, Physica C 235-240 (1994)
  2671--2672.

\bibitem{Maletta}
H.~Maletta, A.~P. Malozemoff, D.~C. Cronemeyer, C.~C. Tsuei, R.~L. Greene,
  J.~G. Bednorz, K.~A. M\"uller, Diamagnetic shielding and {M}eissner effect in
  the high {T}c superconductor $\mathrm{Sr_{0.2}La_{1.8}CuO_4}$, Solid State
  Commun. 88 (1993) 837--840.

\bibitem{Decca}
R.~Decca, R.~Isoardi, F.~de~la Cruz, Linear field dependence of flux flow
  resistance in high temperature superconductors, Solid State Commun. 86 (1993)
  103--107.

\bibitem{Shcherbakov}
A.~S. Shcherbakov, V.~E. Startsev, E.~G. Valiulin, A.~N. Petrov, Time
  dependence of mixed state magnetisation in $\mathrm{La_{1.8}Sr_{.2}CuO_4}$,
  J. Phys.: Condens. Matter 2 (1990) 2199--2204.

\bibitem{Leylekian2}
L.~Leylekian, M.~Ocio, J.~Hammann, Gauge glass properties in a granular
  $\mathrm{La_{1.8}Sr_{.2}CuO_4}$ superconductor, Physica B 194--196 (1994)
  1865--1866.

\bibitem{Rodrguez}
E.~Rodr\'iguez, J.~Luzuriaga, C.~A. D'Ovidio, D.~A. Esparza, Softening of the
  flux-line structure in $\mathrm{La_{1.8}Sr_{0.2}CuO_4}$ measured by a
  vibrating reed, Phys. Rev. B 42 (1990) 10796--10799.

\bibitem{Levy2}
P.~Levy, H.~Ferrari, V.~Bekeris, C.~Acha, A{C} susceptibility and local
  intergranular magnetic field in high-{T}$_c$ superconductors, Physica C 214
  111--118.

\bibitem{Tanaka9}
I.~Tanaka, H.~Kojima, Superconducting single crystals, Nature 337 (1989) 21.

\bibitem{Takagi10}
H.~Takagi, T.~Ido, S.~Ishibashi, M.~Uota, S.~Uchida,
  Superconductor-to-nonsuperconductor transition in
  $\mathrm{(La_{1-x}Sr_x)_2CuO_4}$ as investigated by transport and magnetic
  measurements, Phys. Rev. B 40 (1989) 2254--2261.

\bibitem{GLM}
V.~Ginzburg, L.~Landau, Zh. Eksp. Teor. Fiz. 20 (1954) 1064, english
  translation in \textit{Men of physics: L.D. Landau}, Pergamon (1965).

\bibitem{VortexState00}
J.~R. Clem, The {V}ortex {S}tate, Vol. 438 of NATO ASI series C: Mathematical
  and Physical Sciences, Kluwer Academic Publishers, Ch. Anisotropic
  Superconductors: Fundamentals of Vortices in Layered Superconductors, pp.
  25--39.

\bibitem{HuPRB6}
C.-R. Hu, Numerical {C}onstants for {I}solated {V}ortices in {S}uperconductors,
  Phys. Rev. B 6 (1972) 1756--1760.

\bibitem{Klemm00}
R.~Klemm, J.~Clem, Lower critical field of an anisotropic type-{II}
  superconductor, Phys. Rev. B 21 (1980) 1868.

\bibitem{Zaleski00}
A.~Zaleski, J.~Klamut, Anisotropy of the penetration depth in
  $\mathrm{La_{2-x}Sr_xCuO_4}$ in underdoped and overdoped regions, J. Phys.:
  Condens Matter 11~(48) (1999) 9731.

\bibitem{Bugoslavsky2}
Y.~V. Bugoslavsky, A.~L. Ivanov, A.~A. Minakov, S.~I. Vasyurin, Fishtails and
  anisotropy in underdoped {L}a{S}r{C}u{O} single crystal, Physica C 233 (1994)
  67--76.

\bibitem{ConcEncyP99}
J.~Evetts, A.~Campbell, Concise {E}ncyclopedia of {M}agnetic \&
  {S}uperconducting {M}aterials, Pergamon, 1992, Ch. Critical State Model,
  p.~99.

\end{thebibliography}

\end{document}